\documentclass[a4paper,twocolumn,aps,prl,superscriptaddress]{revtex4}
\usepackage{amsmath}
\usepackage{graphicx}
\usepackage{amssymb}

\newcommand{\ket}[1]{\left|#1\right>}
\newcommand{\bra}[1]{\left<#1\right|}

\begin{document}

\title{Optical pumping into many-body entanglement}

\author{Jaeyoon Cho}
\affiliation{QOLS, Blackett Laboratory, Imperial College London, London SW7 2BW, UK}
\author{Sougato Bose}
\affiliation{Department of Physics and Astronomy, University College London, Gower St., London WC1E 6BT, United Kingdom}
\author{M. S. Kim}
\affiliation{QOLS, Blackett Laboratory, Imperial College London, London SW7 2BW, UK}

\date{\today}

\begin{abstract}
We propose a scheme of optical pumping by which a system of atoms coupled to harmonic oscillators is driven to an entangled steady state through the atomic spontaneous emission. It is shown that the optical pumping can be tailored so that the many-body atomic state asymptotically reaches an arbitrary stabilizer state regardless of the initial state. The proposed scheme can be suited to various physical systems. In particular, the ion-trap realization is well within current technology.
\end{abstract}

\maketitle

Generating entanglement is one of the landmarks of modern physics, which has provoked a huge body of work. Entanglement---quantum correlation---is distinguished from the classical counterpart by its coherent nature \cite{horodecki09}. While the correlation itself can be naturally built up by letting multiple bodies mutually interact, the difficulty usually arises in retaining the coherence, since opening an interaction channel inevitably brings in adversarial environmental effects as well. Consequently, both the theoretical and experimental efforts have been devoted mostly to finding a way to isolate the system, whereby the problem reduces to identifying proper time points at which the unitary dynamics gives a useful form of entanglement \cite{haffner05}. This requires preparation of definite initial states and precise control of the system parameters and timing. In addition, the entanglement generation can be aided by measurements \cite{walther05}.

A conceptually different approach is to tailor an open quantum system in such a way that its irreversible dynamics drives the system into a steady state that is entangled~\cite{schneider02,clark03,kraus04,paternostro04,diehl:2008a,verstraete:2009a}. This approach has apparent advantages over those relying on the unitary evolutions. Firstly, the generated entangled state is robust against decoherence because the intrinsic decoherence channel of the system itself is exploited as a resource. Secondly, the steady state is determined by the relationship (e.g., ratio) between different parameters, not by their exact values or temporal profiles, and moreover is independent of the initial state. This largely alleviates the otherwise necessary manipulation and control mentioned above and also provides extra robustness against fluctuations of the parameters (e.g., of laser fields). In some sense, such a possibility is recognizable from the fact that the environmental effects in ordinary quantum optical systems are well accounted for in terms of the zero-temperature bath \cite{carmichael93}. This means that such an environment always plays the role of an entropy sink that renders the system perfectly coherent. For single atoms, optical pumping---more specifically, coherent population trapping---indeed uses this environmental decay, i.e., the spontaneous emission, to prepare an arbitrary pure superposition of two ground states \cite{gray78}. Generalizing this idea to multiple atoms, however, seems daunting and somewhat counterintuitive at first sight because the spontaneous emission occurs independently in each atom, destroying both the correlation and coherence between the atoms. 

In this paper, we show that the concept of optical pumping can indeed be applied to the cases of many atoms when they are coupled to discrete quantum systems, exemplified here by harmonic oscillators. To be specific, we show that as far as the geometry permits, an arbitrary stabilizer state (a broad class of many-body entangled states) of the atoms can be generated as a steady state through the atomic spontaneous emission. This contrasts with all previous works, mostly with two atoms, which have rather been case studies simply revealing that competition between a continuous driving and the environmental decay can lead to a particular entangled (but far from maximally entangled) steady state \cite{schneider02,clark03} or have shown that atoms driven by entangled fields can have a steady-state entanglement \cite{kraus04,paternostro04}. Our systematic study, on the other hand, suggests a clear-cut mechanism to generate a broad class of entangled states with a high fidelity using classical pumping fields. Recent studies offer formalisms for engineering system-environment couplings to obtain a desired many-body entanglement as a steady state \cite{diehl:2008a,verstraete:2009a}. However, as the resulting Lindblad operators are multipartite ones, which require the interaction between the system and the bath to be effectively many-body, the formalism does not necessarily make its physical realization apparent, except for some special instances leading to only a restricted set of entangled states \cite{diehl:2008a} or those aided in large part by unitary manipulations~\cite{weimer:2010a}. Our scheme, on the other hand, deals with realistic physical models: we show that our idea is directly applicable to ion-trap systems and also to many other systems by allowing only single-atom unitary transformations in addition. Thanks to these exclusive features, our scheme would find straightforward applications, e.g., in establishing a robust entanglement channel between computational nodes or continuously stabilizing an entangled state generated by a different means. As highlighted in recent literatures, our scheme would also have implications in the context of quantum many-body simulation and quantum computation \cite{diehl:2008a,verstraete:2009a, weimer:2010a,cho:2009a}. 
%In a similar context, many-body entanglement could be generated by directly cooling strongly-correlated atomic systems, which is, however, experimentally more demanding in that the strongly-correlated Hamiltonian should also be realized~\cite{cho:2009a}.

Stabilizer states represent a large set of entangled states including error-correcting codes and graph states \cite{got96,raussendorf01}. Let us employ the following convention. Consider an $N$-qubit system with a $2^{N}$-dimensional Hilbert space $\mathcal{H}$ and $M$ stabilizers $\mathcal{S}_{\mu}\equiv\bigotimes_{j=1}^{N}u_{\mu j}$ with $1\le\mu\le M$, where $u_{\mu j}\in\{\sigma_{j}^{X},\sigma_{j}^{Y},\sigma_{j}^{Z},I_{j}\}$ is one of the Pauli operators or the identity operator acting on the $j$-th qubit. As $[\mathcal{S}_{\mu},\mathcal{S}_{\nu}]=0$ for every pair and $(\mathcal{S}_{\mu})^{2}=\bigotimes_{j=1}^{N}I_{j}$, the Hilbert space can be divided into $2^{M}$ subspaces, each with dimension $2^{N-M}$, according to the eigenvalues of the stabilizers: $\mathcal{H}=\bigoplus_{s_{1},...,s_{M}=\pm1}\mathcal{H}(s_{1},...,s_{M})$, where $\mathcal{S}_{\mu}\ket{\Psi}=s_{\mu}\ket{\Psi}$ for $\ket{\Psi}\in\mathcal{H}(s_{1},...,s_{M})$. Our aim is to pump all the population into the subspace $\mathcal{H}(+1,+1,...,+1)$. If M=N, the dimension of this Hilbert space is 1, i.e., it represents a single pure state, but we can also consider more general cases where $M<N$, in which case the steady state is a certain mixed state in an encoded subspace rather than a pure state.

The stabilizer states that can be prepared in our scheme depend on the geometry of the arrangement of harmonic oscillators and atoms. For simplicity, let us consider a case where one harmonic oscillator is coupled to the atoms. Generalizing this setting to other cases with more harmonic oscillators is straightforward. The underlying idea is to have the atoms collectively coupled to harmonic oscillators, thereby splitting the otherwise degenerate energy levels in the excited manifold of the system according to, in our case, the parity. We then resolve the energy splittings so as to optically pump the population with a particular parity into the disjoint subspace. As will become apparent, such an optical pumping is suitable for the generation of stabilizer states.

\begin{figure}
\includegraphics[width=0.62\columnwidth]{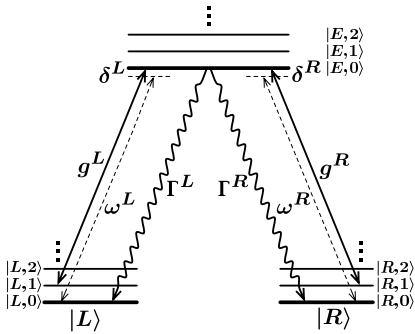}
\caption{Involved atomic levels and transitions. $g^{L,R}$ denotes the coupling rate between the atom and the harmonic oscillator, $\omega^{L,R}$ the Rabi frequency of the pumping field, $\delta^{L,R}$ the detuning, and $\Gamma^{L,R}$ the spontaneous emission rate.}
\label{fig:level}
\end{figure}

We use three-level atoms with two ground levels $\ket{L}$ and $\ket{R}$ and an excited level $\ket{E}$, as shown in \figurename~\ref{fig:level}. The two atomic transitions $\ket{L}\leftrightarrow\ket{E}$ and $\ket{R}\leftrightarrow\ket{E}$ are coupled respectively to orthogonal polarizations of light. Let us denote by $a$ ($a^{\dagger}$) the annihilation (creation) operator for the harmonic oscillator. We assume that the harmonic oscillator interacts with the atom in such a way that they exchange energy, which can be described in the interaction picture by the following Hamiltonian:
\begin{equation}
H_{\text{ah}}=\sum_{j=1}^{N}a\left(g_{j}^{L}\ket{E}_{j}\bra{L}+g_{j}^{R}\ket{E}_{j}\bra{R}\right)+\text{H.c.},
\end{equation}
where the subscript $j$ represents the $j$-th atom. The coupling rates $g_{j}^{L}$ and $g_{j}^{R}$ are assumed to be independently adjustable. On top of this coupling, we also apply classical pumping fields with Rabi frequencies $\omega_{j}^{L}$ and $\omega_{j}^{R}$ and detunings $\delta^{L}$ and $\delta^{R}$, respectively, as in \figurename~\ref{fig:level}. For convenience, let us call the former an $L$-field and the latter an $R$-field. This driving can be described in the interaction picture by the following Hamiltonian:
\begin{equation}
H_{\text{p}}=\sum_{j=1}^{N}\omega_{j}^{L}e^{i\delta^{L}t}\ket{E}_{j}\bra{L}+\omega_{j}^{R}e^{i\delta^{R}t}\ket{E}_{j}\bra{R}+\text{H.c.}
\end{equation}
Once the atom is excited, spontaneous emission to $\ket{L}$ and $\ket{R}$ takes place with rates $\Gamma^{L}$ and $\Gamma^{R}$, respectively, which can be described by the following master equation:
\begin{equation}
\frac{d}{dt}\rho=i[\rho,H_{\text{ah}}+H_{\text{p}}]+\sum_{j=1}^{N}\sum_{x=L,R}\mathcal{L}_{j}^{x}(\rho),
\label{eq:master}
\end{equation}
where $\rho$ is the density matrix of the system and $\mathcal{L}_{j}^{x}(\rho)=\Gamma^{x}\bigl(\ket{x}_{j}\bra{E}\rho\ket{E}_{j}\bra{x}-\frac12\ket{E}_{j}\bra{E}\rho-\frac12\rho\ket{E}_{j}\bra{E}\bigr)$.

\begin{figure}
\includegraphics[width=0.94\columnwidth]{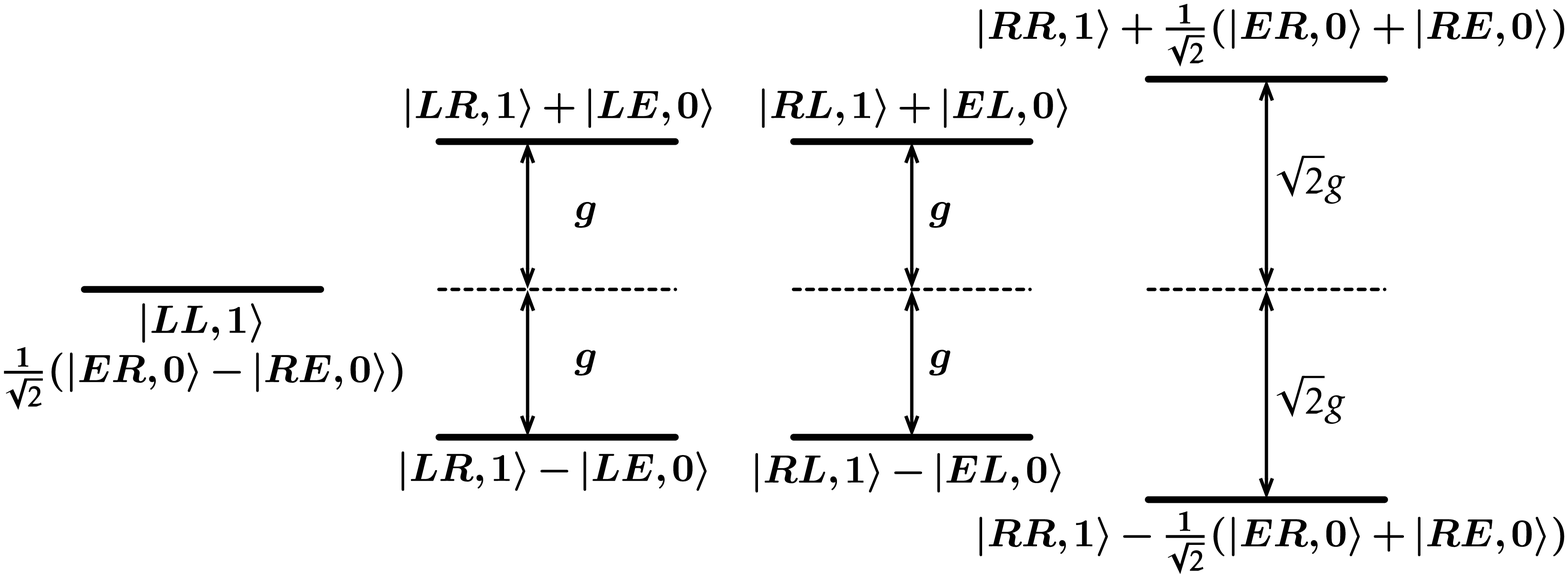}
\centerline{(a)}\\[1em]
\includegraphics[width=0.84\columnwidth]{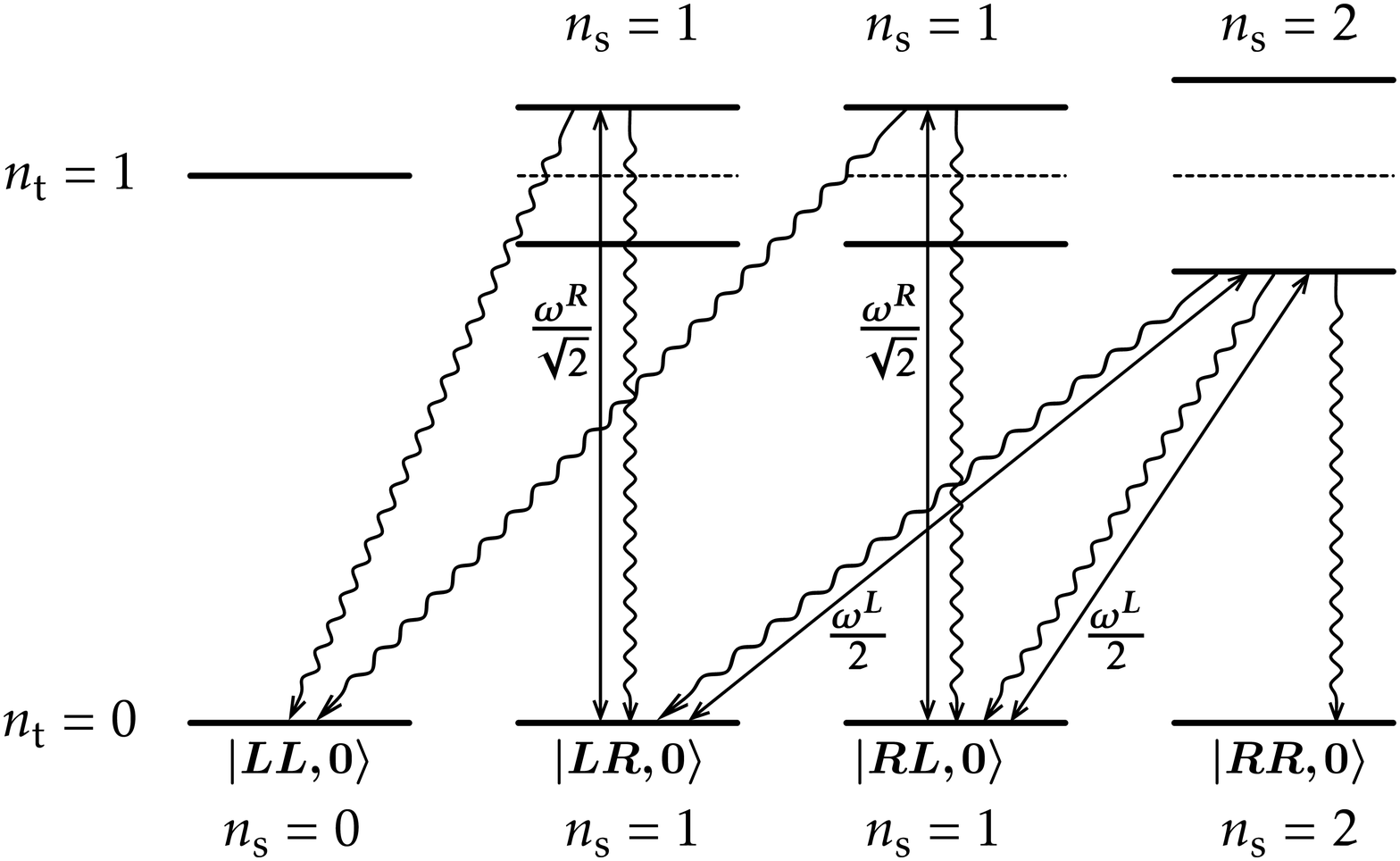}
\centerline{(b)}
\caption{(a) Energy splitting in the manifold of one excitation for two atoms (normalization is ignored). (b) Optical pumping from the odd-parity subspace to the even-parity subspace. Straight arrows represent classical pumping fields and wiggly arrows spontaneous emissions. ${\omega^{r}}/{\sqrt{2}}$ and ${\omega^{l}}/{2}$ are the corresponding effective Rabi frequencies.}
\label{fig:pumping}
\end{figure}

To begin with, suppose only two atoms are coupled to the harmonic oscillator and only the transition $\ket{R}\leftrightarrow\ket{E}$ is coupled, i.e., $g_{1}^{L}=g_{2}^{L}=0$ and $g_{1}^{R}=g_{2}^{R}=g$. We can denote the state by $\ket{\phi,n}$, where $\ket{\phi}$ denotes the atomic state and $\ket{n}$ the $n$-th excited state of the harmonic oscillator. Note that the coupling between the atom and the harmonic oscillator does not change the number of total excitations $n_{\text{t}}\equiv a^{\dagger}a+\sum_{j}\ket{E}_{j}\bra{E}$, i.e., $[H_{\text{ah}},n_{\text{t}}]=0$. Putting the pumping fields and the spontaneous emission aside, the Hilbert space is thus divided into mutually decoupled subspaces according to $n_{\text{t}}$. Furthermore, due to the coupling, the degeneracy in each excited manifold is lifted. In the manifold of $n_{\text{t}}=1$, transitions $\ket{LR,1}\leftrightarrow\ket{LE,0}$ and $\ket{RL,1}\leftrightarrow\ket{EL,0}$ occur with coupling rate $g$, whereas transition $\ket{RR,1}\leftrightarrow\frac{1}{\sqrt{2}}(\ket{ER,0}+\ket{RE,0})$ occurs with coupling rate $\sqrt{2}g$ as the atoms collectively interact with the harmonic oscillator. Consequently, as shown in \figurename~\ref{fig:pumping}(a), the excited manifold shows energy splittings, whose amounts are determined by the number of atoms in states $\ket{R}$ and $\ket{E}$. Formally, one can define $n_{\text{s}}=\sum_{j}\ket{R}_{j}\bra{R}+\ket{E}_{j}\bra{E}$ and then the energy splitting is given by $\pm\sqrt{n_{\text{s}}}g$ (except for uncoupled states). 

We resolve these energy splittings by adjusting the detunings to achieve selective pumping of the population. As we aim to generate an entangled state in the ground-state manifold (i.e., $n_{\text{t}}=0$) and the spontaneous emission always projects the population to the lower manifold, we need to consider only the two lowest manifolds of $n_{\text{t}}=0$ and $n_{\text{t}}=1$. \figurename~\ref{fig:pumping}(b) depicts the pumping process from the subspace with $\sigma_{1}^{Z}\sigma_{2}^{Z}=-1$ to that with $\sigma_{1}^{Z}\sigma_{2}^{Z}=+1$. Here, the Pauli operators are defined over the ground levels of the atoms, e.g., $\sigma_{j}^{Z}=\ket{L}_{j}\bra{L}-\ket{R}_{j}\bra{R}$. Note that the $L$- and $R$-fields couple the two manifolds in different ways. While the $L$-field induces transition in such a way that when the state is excited $n_{\text{s}}$ is increased by one, the $R$-field preserves $n_{\text{s}}$. The spontaneous emission, on the other hand, only decreases $n_{\text{t}}$ by one while either decreasing $n_{\text{s}}$ by one or preserving $n_{\text{s}}$ (see \figurename~\ref{fig:pumping}(b)). The overall effect of the pumping is thus that the $L$-field increases $n_{\text{s}}$ by one while the $R$-field decreases $n_{\text{s}}$ by one. Thanks to the energy splittings, one can apply these optical pumpings selectively to the population with a particular $n_{\text{s}}$: in order to pump the population with $n_{\text{s}}$ to the subspace with $n_{\text{s}}+1$ ($n_{\text{s}}-1$), we apply an $L$-field ($R$-field) to every atom with detuning $\pm\sqrt{n_{\text{s}}+1}g$ ($\pm\sqrt{n_{\text{s}}}g$).

It is easily seen that the above pumping process drives a two-qubit system into the subspace with $\mathcal{S_{\mu}}=\sigma_{1}^{Z}\sigma_{2}^{Z}=+1$. Generalizing this idea to the case of an arbitrary $\mathcal{S_{\mu}}$ is straightforward. Suppose first that we need to pump the population with $\sigma_{1}^{Z}\sigma_{2}^{Z}\cdots\sigma_{N}^{Z}=-1$ to the subspace with $\sigma_{1}^{Z}\sigma_{2}^{Z}\cdots\sigma_{N}^{Z}=+1$. We achieve this by pumping every population with an odd $n_{\text{s}}$, i.e., that with an odd $(N+\sigma_{1}^{Z}+\cdots+\sigma_{N}^{Z})/2$, to the subspace with $n_{\text{s}}\pm1$. For this, we apply to every atom $L$-fields with detunings $\sqrt{2}g$, $\sqrt{4}g$, $\sqrt{6}g$, and so forth. In the same fashion, we apply to every atom $R$-fields with detunings $-g$, $-\sqrt{3}g$, $-\sqrt{5}g$, and so forth. Here, we exploit red (blue) detunings for $L$-fields ($R$-fields), but this is not mandatory. For a general $\mathcal{S}_{\mu}$, we can achieve the same kind of optical pumping by rotating the basis of the operation for each $j$-th atom according to $u_{\mu j}$. If $u_{\mu j}=I_{j}$, we turn off all the couplings and pumping fields for the $j$-th atom so that the atom is excluded from the pumping process. Otherwise, we adjust the coupling rates so that $g_{j}^{L}=\alpha_{j}^{*}g$ and $g_{j}^{R}=\beta_{j}^{*}g$, where $\alpha_{j}\ket{L}_{j}+\beta_{j}\ket{R}_{j}$ is the $+1$ eigenstate of $u_{\mu j}$, and instead of an $R$-field with Rabi frequency $\omega_{j}^{R}$, we apply an $L$-field with $\alpha_{j}^{*}\omega_{j}^{R}$ and an $R$-field with $\beta_{j}^{*}\omega_{j}^{R}$ \cite{note1}. Likewise, instead of an $L$-field with Rabi frequency $\omega_{j}^{L}$, we apply an $L$-field with $\beta_{j}\omega_{j}^{L}$ and an $R$-field with $-\alpha_{j}\omega_{j}^{L}$. 

A general stabilizer state is now generated by performing the optical pumping processes for all $\mathcal{S}_{\mu}$'s one by one as described above. An important question here is how the performance, namely, the characteristic time for the system to reach the steady state, scales as the number of atoms increases. Unfortunately, it turns out that the time increases exponentially with the number of stabilizers $M$. In order to see this, it is convenient to think of the pumping process as a random walk (or hopping) among the $2^{M}$ subspaces $\mathcal{H}(s_{1},...,s_{M})$. For example, suppose a linear cluster state $\ket{\Psi_{N}}=2^{-N/2}\bigotimes_{j=1}^{N-1}(\ket{L}_{j}\bra{L}+\ket{R}_{j}\bra{R}\sigma_{j+1}^{Z})\bigotimes_{k=1}^{N}(\ket{L}_{k}+\ket{R}_{k})$ of four atoms ($N=4$), stabilized by $\mathcal{S}_{1}=\sigma_{1}^{X}\sigma_{2}^{Z}I_{3}I_{4}$, $\mathcal{S}_{2}=\sigma_{1}^{Z}\sigma_{2}^{X}\sigma_{3}^{Z}I_{4}$, $\mathcal{S}_{3}=I_{1}\sigma_{2}^{Z}\sigma_{3}^{X}\sigma_{4}^{Z}$, and $\mathcal{S}_{4}=I_{1}I_{2}\sigma_{3}^{Z}\sigma_{4}^{X}$, and consider the pumping step for $\mathcal{S}_{1}$. The original aim of this step is to pump the population in $\mathcal{H}(-1,s_{2},s_{3},s_{4})$ to $\mathcal{H}(+1,s_{2},s_{3},s_{4})$. However, as it destroys the coherence of atom 1 and atom 2, the values of $\mathcal{S}_{2}$ and $\mathcal{S}_{3}$, which contain a Pauli operator on atom 1 or atom 2, are also affected. Consequently, the pumping for $\mathcal{S}_{1}$ in fact occurs from $\mathcal{H}(-1,s_{2},s_{3},s_{4})$ to one of $\mathcal{H}(+1,\pm1,\pm1,s_{4})$ randomly. The performance is then assessed by estimating the average number of steps to reach the subspace $\mathcal{H}(+1,...,+1)$ by such a random walk, along with the actual pumping time for each step. The latter increases as the number of Pauli operators in $\mathcal{S}_{\mu}$, which we denote by $k_{\mu}$, increases because the energy gap to resolve $g(\sqrt{k_{\mu}}-\sqrt{k_{\mu}-1})$ decreases. This scales only polynomially and can be bounded as long as $k_{\mu}$ is finite for all $\mathcal{S}_{\mu}$, which is usually the case. However, the former, the number of steps to reach the steady state, increases exponentially as $\lesssim M(2^{l_{\text{m}}})^{M}$, where each pumping step affects at most $l_{\text{m}}$ other stabilizers. Our scheme is thus more relevant for a moderate number of atoms. 

\begin{figure}
\begin{tabular}{ccc}
\includegraphics[width=0.44\columnwidth]{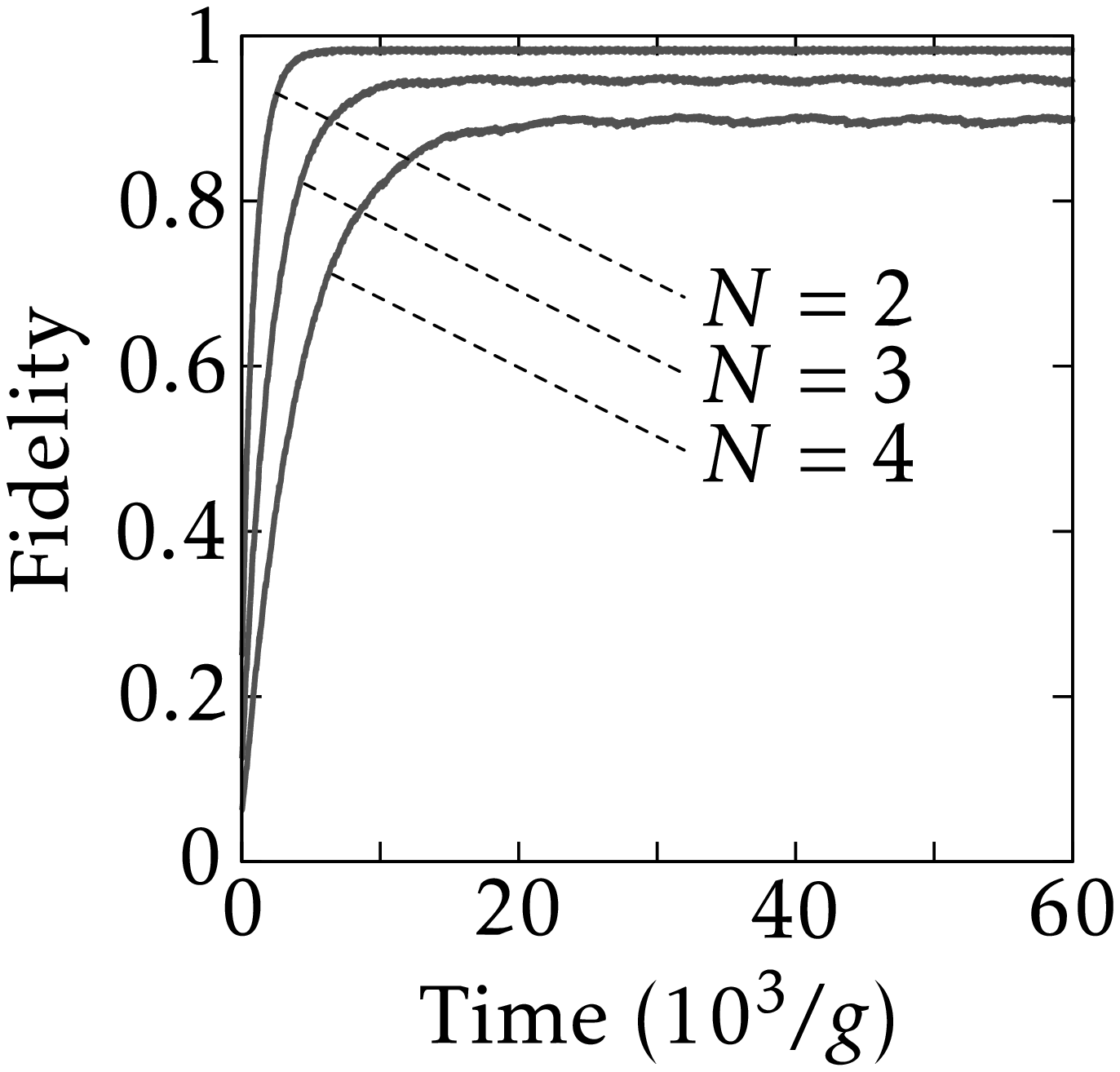} &
\hspace{1em} &
\includegraphics[width=0.44\columnwidth]{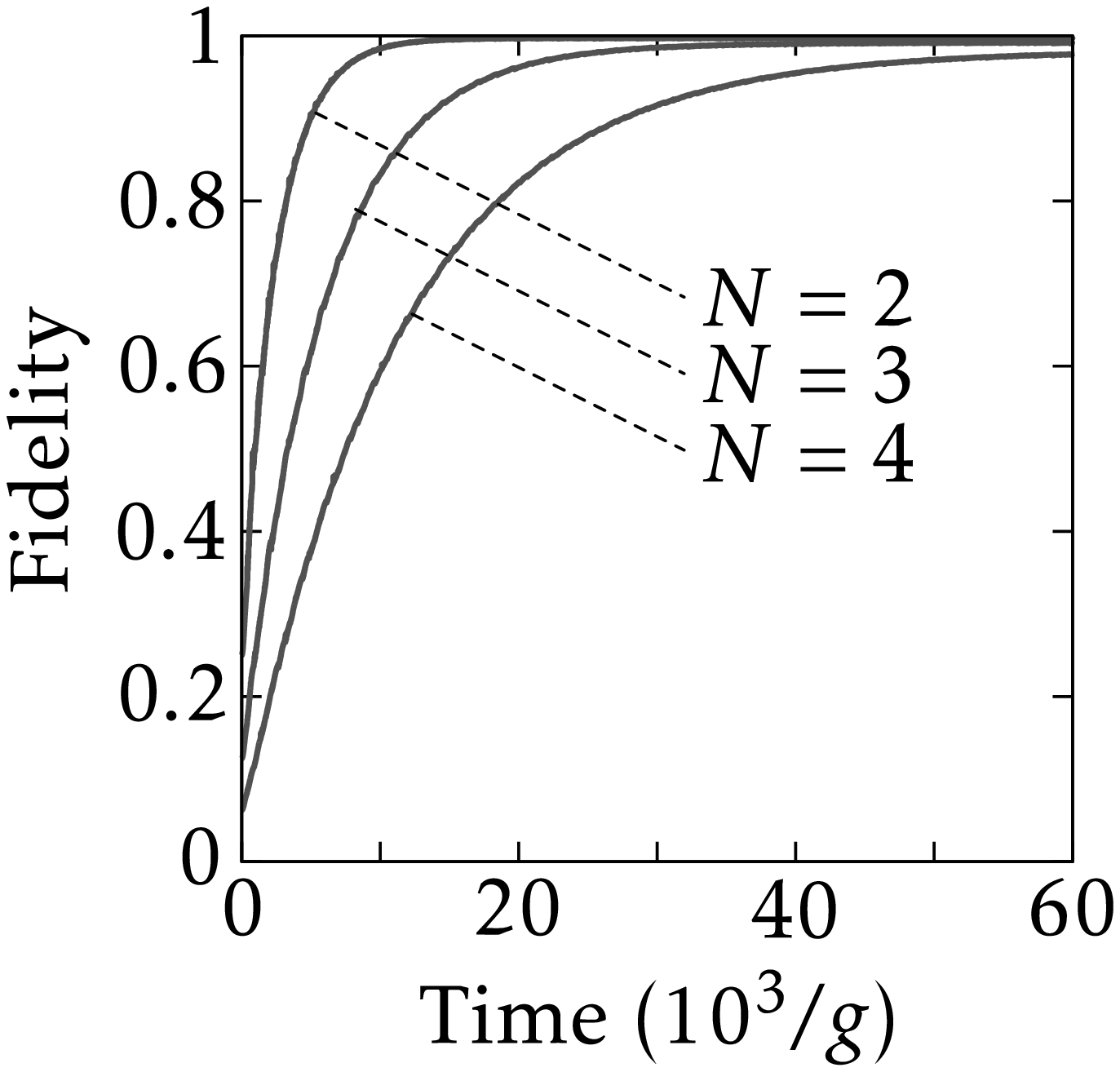} \\
(a) $\lambda=0.05$ & & (b) $\lambda=0.02$
\end{tabular}
\caption{Optical pumping to linear cluster states for $N=2,3,4$ when a fully mixed state is taken as the initial state for (a) $\lambda=0.05$ and (b) $\lambda=0.02$. $\lambda$ is a constant determining the degree to which the energy splitting is resolved (see text).}
\label{fig:graph}
\end{figure}

\figurename~\ref{fig:graph} shows the results of our simulation for the generation of linear cluster states, which is done by numerically integrating the master equation \eqref{eq:master}. As the pumping exploits the energy splitting, both the pumping field and the spontaneous emission should be weak enough compared to the energy gap so that the energy splittings can be resolved. For this, we choose the parameters as $\Gamma=\lambda g(\sqrt{k_{\mu}}-\sqrt{k_{\mu}-1})$, $\Gamma^{L}=\Gamma^{R}=\Gamma/2$, and $\omega^{L}=\omega^{R}=\Gamma/\sqrt{k_{\mu}}$, where $\lambda\ll1$ is a small constant, and each round of the pumping for one stabilizer is performed for a period of time $\pi/\Gamma$. Our particular choice of the parameters is, however, not mandatory as the pumping is by nature insensitive to the exact values of the parameters. Taking a fully mixed state of atoms along with the ground state of the harmonic oscillator as an initial state, \figurename~\ref{fig:graph} shows the time evolution of the fidelity of the state to the desired linear cluster state. The weaker the pumping field and the spontaneous emission are (smaller $\lambda$), the higher is the final fidelity, but the longer it takes to reach the steady state. One could vary the parameters in time to optimize the performance, e.g., by taking larger $\lambda$ during the initial transient period and taking smaller $\lambda$ afterward to obtain a higher fidelity. 

It is evident that the present scheme is well suited to the existing ion-trap systems used for quantum information processing \cite{schmidt03,leibfried03}. For example, $\text{Ca}^{+}$ ions can be used, where the two hyperfine levels of $S_{1/2}$ represent $\ket{L}$ and $\ket{R}$, respectively, and one of $D_{5/2}$ represents $\ket{E}$ \cite{roos99}. The coupling between each ion and the center of mass mode of the ions $g_{j}^{L,R}$ can be controlled individually by focused beams. As $D_{5/2}$ is metastable, the spontaneous emission rate $\Gamma^{L,R}$ can be controlled by adjusting pumping fields via $P_{3/2}$ (and repumping fields via $P_{1/2}$ to make the transition closed). As the typical coupling rate is $g\gtrsim100~\text{kHz}$, the system reaches the steady state in a few hundred milliseconds for a small number of atoms, as can be inferred from \figurename~\ref{fig:graph}. One can also incorporate sympathetic cooling into the pumping sequence to overcome motional heating \cite{larson86}.

The present scheme with slight modifications can be applied to many other physical systems. First, we can consider a case where only one of the two atomic transitions is persistently coupled, i.e., $g^{L}=0$ and $g^{R}\not=0$. This enables the optical pumping for stabilizer $\bigotimes_{j}\sigma_{j}^{Z}$. For different stabilizers, one performs single-qubit rotations to the atoms corresponding to $\sigma_{j}^{X,Y}$ before and after each pumping and detune the transition by an ac Stark shift for those corresponding to $I_{j}$. It is easily seen that this modified scheme is well suited to systems of atoms sitting in a common cavity mode. Another modification can be made by noting that the harmonic oscillator can in fact be replaced by a two-level atom, where the interaction between atoms is the spin-exchange interaction. Along with single-qubit manipulations as described above, which are relatively easy to do, our scheme can thus be applied to various Heisenberg spin systems realized with atoms, where it is understood that auxiliary levels are needed to  represent another ground level $\ket{L}$ and to realize spontaneous emissions effectively if required \cite{micheli06,angelakis07}.

We thank Ho Tsang Ng and Sun Kyung Lee for fruitful discussions. We acknowledge the UK EPSRC for financial support.


\begin{thebibliography}{22}
\bibitem[{not()}]{note1}
\bibinfo{note}{Here, the coupling rates and the Rabi frequencies are complex
  numbers.}
\end{thebibliography}
\end{document}